\documentclass[journal]{IEEEtran}
\usepackage[utf8]{inputenc}
\usepackage[T1]{fontenc}

\usepackage{graphicx}
\usepackage{subfig}
\usepackage{float}
\usepackage{url}
\usepackage{xspace}
\usepackage[super]{nth}
\usepackage[noadjust]{cite}
\usepackage[shortlabels, inline]{enumitem}

\newcommand{\eg}{\textit{e.g.,}\xspace} 
\newcommand{\ie}{\textit{i.e.,}\xspace}

\usepackage{color, xcolor, colortbl}
\newcommand{\boubakr}{\color{black}}
\newcommand{\afaf}{\color{black}}

\newcommand{\rev}{\color{black}}

\usepackage{tabularx}
\newcolumntype{P}[1]{>{\centering\arraybackslash}p{#1}}
\newcolumntype{M}[1]{>{\centering\arraybackslash}m{#1}}
\newcolumntype{L}[1]{>{\centering\arraybackslash}l{#1}}
\newcolumntype{Z}{>{\centering\let\newline\\\arraybackslash\hspace{0pt}}X}

\newcolumntype{Y}{>{\raggedright\let\newline\\\arraybackslash\hspace{0pt}}X}

\begin{document}
\bstctlcite{IEEEexample:BSTcontrol}
%TC:ignore
\title{Empowering Prosumer Communities in Smart Grid with Wireless Communications and Federated Edge Learning}

\author{
    \IEEEauthorblockN{
        Afaf Ta\"ik, \IEEEmembership{Student Member, IEEE},
        Boubakr Nour, \IEEEmembership{Member, IEEE}, and
		Soumaya Cherkaoui, \IEEEmembership{Senior Member, IEEE}
    }
    
    \thanks{A. Ta\"ik, B. Nour, and S. Cherkaoui are with the INTERLAB Research Laboratory, Faculty of Engineering, Department of Electrical and Computer Science Engineering, Université of Sherbrooke, Sherbrooke (QC) J1K 2R1, Canada (emails: afaf.taik@usherbrooke.ca, boubakr.nour@usherbrooke.ca, soumaya.cherkaoui@usherbrooke.ca).}
}

\markboth{}{} % IEEE Wireless Communications

\maketitle

\begin{abstract}
    The exponential growth of distributed energy resources is enabling the transformation of traditional consumers in the smart grid into prosumers. Such transition presents a promising opportunity for sustainable energy trading. Yet, the integration of prosumers in the energy market imposes new considerations in designing unified and sustainable frameworks for efficient use of the power and communication infrastructure. Furthermore, several issues need to be tackled to adequately promote the adoption of decentralized renewable-oriented systems, such as communication overhead, data privacy, scalability, and sustainability.
    In this article, we present the different aspects and challenges to be addressed for building efficient energy trading markets in relation to communication and smart decision-making. Accordingly, we propose a multi-level pro-decision framework for prosumer communities to achieve collective goals.  Since the individual decisions of prosumers are mainly driven by individual self-sufficiency goals, the framework prioritizes the individual prosumers' decisions and relies on the 5G wireless network for fast coordination among community members. In fact, each prosumer predicts energy production and consumption to make proactive trading decisions as a response to collective-level requests. Moreover, the collaboration of the community is further extended by including the collaborative training of prediction models using Federated Learning, assisted by edge servers and prosumer home-area equipment. In addition to preserving prosumers' privacy, we show through evaluations that training prediction models using Federated Learning yields high accuracy for different energy resources while reducing the communication overhead.
\end{abstract}

%TC:endignore
\IEEEpeerreviewmaketitle

\section{Introduction}
\label{sec:introduction}
%% The power grid has become unable to cope with the increasing energy demand. In fact, the existing system is centralized and mainly dependent on non-renewable energy sources. Moreover, the business model consists of the simple interaction between distribution and transmission, with no active consumer involvement. The necessity for a dynamic smart grid is motivated by the shift from conventional energy sources towards renewables. Such dynamicity, in addition to demand-response systems, lead to the emergence of prosumers (i.e., producers and consumers) who are capable of both generating and consuming energy. 

%%The traditional power grid was designed to meet user demands (energy consumption) and to provide real-time load management, where the conventional business model consists of simple interactions between distribution and transmission systems, without active consumer participation. With the changing nature of grid operations, the dynamics of demand response, the large-scale integration of renewable sources, and the growing need for consumer interaction, the use of a more flexible system, called the smart grid, has become necessary~\cite{looney2020statistical}. 

The smart grid is an evolved power system with an integrated two-way flow of energy and information whose purpose is to enhance the efficiency of the system. The ubiquity of wireless communications has broadly promoted the deployment of sustainable power systems and helped monitor and control various operations of smart grids {\boubakr \cite{rehmani2018integrating}}. The higher performance and improved efficiency of fifth-generation (5G) communication networks are expected to reinforce this trend by delivering the ultra-low latency, the massive connectivity, and ultra-reliability required by smart grid services. 

Communication technologies have not only changed the nature of the power system in terms of monitoring and operating processes, but they have also enabled the emergence of new behaviors and market models. In particular, the emergence of prosumers {\boubakr (\ie users that can produce and consume energy)} is one of the strongest trends in the field of renewable energy in the smart grid. {\rev Multiple prosumers can now collectively create a {\em Prosumer Community Group} (PCG). PCG aims to producing and sharing energy in large amounts between users and utilities, providing a unified platform for information trading among neighbors within the local community, as well as interfacing with external prosumers and other energy entities. Indeed, PCG can effectively supply distributed generation, storage, and demand response~\cite{aloqaily2020energy}.}
Nevertheless, {\boubakr with the new opportunities offered by the advent of prosumers, unprecedented challenges are emerging in the management of energy production and demand in the grid.} Indeed, the relationship between energy consumption and production is not always equiponderant. Hence, many techniques, including machine learning (ML), have been integrated to predict energy consumption, and adjust the available capacity accordingly. With the emergence of PCGs, there is also a need to forecasting energy production, which is becoming as important as predicting energy consumption, in order to adjust available resources effectively.

{\boubakr Current prediction systems in smart grids are based on collecting demand/production information and analyzing data at cloud servers, where the value chain relies on access to data}. In 5G enabled smart grids, it is expected that prosumer systems will involve massive machine type communication (mMTC) where downlink communication is used for power control and uplink communication is used for data collection~\cite{GSMA:2020:Online}. However, this data collection model results in considerably large datasets to be fed into the prediction system. More importantly, in many jurisdictions, privacy and data protection legislation requires the development of new operating models in which there is no personal data collection. %~\cite{bessa2018data}

Federated Learning (FL)~\cite{konevcny2016federated}, an extension of machine learning, enables multiple devices/servers to collaboratively learn a prediction model in a distributed manner while keeping all training data locally on the device, thus decoupling machine learning capability from the need to store data in a centralized entity. In Federated Edge Learning (FEEL)~\cite{taik2020federated,taik_data-aware_2021}, the global model training is performed at the Edge of the network. FEEL is particularly useful for delay-sensitive applications or congested backhaul networks, and it has been proposed as an interesting tool for household electrical load forcasting~\cite{taik2020electrical}. It provides a promising paradigm to enable decentralized learning without compromising data privacy. 

The motivation behind this work is not only to design a privacy-aware integration mechanism for prosumers in 5G-enabled smart grid architectures but also to enable prosumers to make optimal decisions using short-term and long-term predictions, as well as allow PCGs to build reliable decision processes collaboratively that maximize the length of the energy production-consumption relationships. Coupling the reliable wireless communications provided by 5G networks with intelligent learning supported by FEEL will help maximize stakeholder's profits in the energy market.

The purpose of this work is twofold:
\begin{itemize}
    \item To enable a multi-level decision process at the aggregator and the individual prosumer levels.
    
    \item To collaboratively build prediction models through federated learning and preserve the eco-system data privacy.
\end{itemize}
{\boubakr Specifically, the contribution of this work can be summarized as follows: 
\begin{enumerate*}[(1)]
    \item we discuss key elements for 5G empowered energy markets and highlight challenges related to their design,
    
    \item we design a multi-stage energy forecasting framework and a decision process for PCGs empowered with FL using edge equipment, and
    
    \item  using real datasets, we evaluate the accuracy of load forecasts and the potential network load gain through simulations. 
\end{enumerate*}}

The remainder of this paper is structured as follows.
% Section II presents an overview of prosumers integration in the smart grid . In Section III, we define the proposed architecture and decision process. Section IV introduces the simulations and numerical results. Section VI concludes the paper.
The next section overviews prosumers' integration in the smart grid along with existing issues and challenges. Then, we present the proposed architecture and detail the decision process framework using FEEL. Later, we discuss the simulation and obtained results. Finally, we conclude the paper and present some future work.

\section{Prosumers in Smart Grid: An Overview}
\label{sec:background}
In this section, we present the key elements for prosumers' integration in smart grids. We focus on the integration models, communication,  and decisions and planning. Then, we present the existing issues and challenges in relation to these elements.

\subsection{Key Elements in 5G Empowered Prosumers Markets}
The increase in population and industrial/commercial sectors drives an unprecedented growth in energy demand. This expansion leads to peak times when most users are simultaneously using electricity. 

%{\em Demand-Response} (DR) has emerged as a solution to control the electrical load. 
%It offers users financial incentives in exchange for shutting down part of peak demands and/or delaying non-time-critical tasks to off-peak hours.
% 
To overcome the electricity shortages, users may contribute to energy production and feed the network when needed. A user that simultaneously produces and consumes energy is known as a {\em prosumer}. A prosumer can produce energy using small renewable energy sources and store it in battery banks or Electric Vehicles (EVs)~\cite{espe2018prosumer,said_scheduling_2014} . In fact, an ever-increasing number of customers have local generation capability (\ie Distributed Energy Resources -- DERs), in addition to several adaptable loads, such as thermostatically-controlled loads and distributed energy storage devices. Moreover, EVs are also appealing as controllable loads \cite{said_advanced_2013,rezgui_smart_2017,said_queuing_2013} because they can be restricted for significant periods of time with no significant impact on end-use function. 

The emergence of prosumers imposes new considerations for seamless smart grid integration. These considerations can be labeled into three levels:
\begin{enumerate*}[(i)]
    \item integration and control model (\eg market model),
    \item communication (\eg requests, coordination), and
    \item pro-active decision making (\eg forecasting).
\end{enumerate*}

%Yet, it works well only when the energy demand is less than the grid's generated energy. Otherwise, consumers may encounter shortages of electricity, which causes severe issues in productivity and transportation.

% \begin{figure*}[!t]
% 	\centering
% 	\includegraphics[width=.75\linewidth]{fig/market_challenges}
% 	\caption{Mapping Smart Grid Key elements to challenges.}
% 	\label{fig:market_challenges}
% \end{figure*}

\vspace{0.2cm}
\textbf{Prosumers Integration Models:}
The increasing number of prosumers imposes significant changes on the electricity market and provides various opportunities for exchanging and balancing electricity production and demand. Future strategies require an efficient integration of prosumers into the competitive electricity market.  In fact, several promising approaches have been proposed, such as peer-to-peer models, indirect customer-to-customer trading, and prosumer community groups.

\begin{itemize}
    \item Peer-to-Peer (P2P): In this model, prosumers are interconnected directly with each other in a completely decentralized manner. While this approach offers high flexibility and total autonomy, it comes with a high communication overhead underlying the search for suitable trading prosumers. Furthermore, an individual prosumer may not be able to produce energy that sufficient to match the fluctuating demand of a peer.

    \item Indirect Trading: Due to the nature of P2P networks, the search for suitable trading prosumers may be challenging and time-consuming. A potential solution to overcome this issue is energy brokers. The latter can be used to create a match between producers and consumers. However, relying on a central entity to manage the trading operations is not a scalable solution, as the broker will be overwhelmed when all the individual prosumers send the trading requests simultaneously.
    
    \item Prosumer Community Groups: On account of the freedom offered by the P2P paradigm and the organizational aspect of energy brokers, a group of prosumers can collectively -- based on their behavior profiles and geographical structures, form a community known as PCG.
    % A PCG provides a unified platform for local energy and information trading among neighbors within the local community, as well as interfacing with external prosumers and other energy entities.
    The goals of a PCG are:
    \begin{enumerate*}[(i)]
        \item achieve a sustainable energy exchange,
        \item fulfill the energy demands for external customers,
        \item increase the income, and
        \item reduce the costs.
    \end{enumerate*}
\end{itemize}

\noindent The integration of various DERs and EVs provides further opportunities for the development of innovative business models and energy ecosystems. However, the heterogeneous nature of energy sources {\afaf \cite{rehmani2018integrating}} and their requirements urge novel approaches for prosumers' market design. Determining the integration model of the prosumers is indispensable to determine the required infrastructure for power distribution and communication.

\vspace{0.2cm}
\textbf{Communication:}
Given the high uncertainties of consumption patterns, modern measurement units support fast monitoring with data refresh up to 50 times per second. Such updated frequencies open up enormous possibilities for fine-grained network control to support the complex energy markets' transition towards more decentralized renewable-oriented systems. Smart grids change the way energy is stored and delivered through vital information sharing among all the connected components. Home Area Network (HAN), smart meters, and home energy management systems (HEMSs) are largely used to prompt behavioral changes in energy consumption and storage, while among individual prosumers and PCGs, 5G networks facilitate information management through, for instance, the broadcast of information (\eg price signal, weather data). Therefore, a reliable communication system is critical for the management of smart grids.

\vspace{0.2cm}
\textbf{Pro-active Decisions and Planning:}
Analyzing and predicting prosumers' behavior profiles are crucial during energy trading planning and future market design. Load forecasting is a key element for proactive decision-making, as it allows to measure the projected energy supply and the future consumption and thus change the operation strategy accordingly. Different horizon forecasts are subject to several studies in smart grids, as each serves a specific purpose according to the length of the forecast duration. For prosumer markets, the following forecast horizons are considered:
% \cite{deligiannis2019predicting}
\begin{itemize}
    \item Very short-term forecasting (seconds to minutes ahead) can be used for storage control for Vehicle-to-Grid (V2G), which are used to adjust small fluctuations. It is also necessary for electricity market clearing. 
    
    \item Short-term forecasting (24 to 72 hours ahead) is crucial for key decision-making problems involved in the electricity market, such as economic dispatch and unit commitment.
    
    \item Medium and long-term forecasting (weeks to years ahead) are useful for maintenance scheduling in future systems and market planning.
\end{itemize}

An overall understanding of the factors that dominate prosumers' behaviors and their interactions within the smart grid enables designing an efficient decision-making framework. While ML techniques enable designing intelligent and responsive energy markets, many challenges still lack extensive studies.

\subsection{Issues and Challenges}
The proliferation of DERs yields many challenges to market design for proactive distribution systems. The limitations of existing work are not essentially based on their working principle; instead, they are more related to security, privacy, with seamless integration into 5G networks.
% 
%Figure~\ref{fig:market_challenges} maps the smart grids market key elements into different challenges and issues based on different aspects.
% 
In the following, we dissect the existing challenges and explore the technical requirements for an optimal solution.

\vspace{0.2cm}
\textbf{Heterogeneous DERs:}
Heterogeneous DERs, such as wind power and photo-voltaic (PV) are affected by different external conditions. For instance, solar radiation and solar altitude have substantial effects on PV, while wind power is directly related to wind speed. Since accurate predictions of these weather factors are challenging to obtain, their uncertainties inevitably lead to large errors when predicting DER production.
Moreover, the topology information is often unavailable due to frequent changes at the individual prosumers level. For instance, EVs can appear once every few hours per day, making it hard for an aggregator to forecast EVs' willingness to share energy. Besides, these individual prosumers are unable to compete with traditional energy generators, as their energy supply is small and often unpredictable. The high fluctuations in the consumption and generation profiles are due to the volatility of the production and consumption profiles, which motivates coordination among prosumers on a shorter time scale and personalized level to attain more accurate results. The coordinated control of heterogeneous DERs can provide further opportunities for achieving better energy regulation.

\vspace{0.2cm}
\textbf{Consumption Load Forecasting:}
While energy production depends heavily on uncertain weather conditions, consumption, on the other hand, is affected by the users' habits and behavior. Several prediction methods (\eg Support Vector Regression, Long Short-Term Memory)~\cite{taik2020electrical,saputra2019energy} have been widely used in the literature to predict consumption profiles. 
Hybrid approaches combining time-series forecasting with other methods, such as decision trees and consumer clustering help improve the forecasting results.
Nonetheless, the individual short-term load profiles remain uncertain and hard to predict compared to an aggregate prediction. Therefore, a PCG selling energy to the grid is more reliable in providing a more sustainable energy supply in contrast to individual prosumers.  

\vspace{0.2cm}
\textbf{Communication Overhead:}
As system operators need to assess the reliability of DERs and measure their participation to maintain the market equilibrium, reliable communication is a crucial enabler for smart energy markets. Moreover, the rapid growth of EVs adoption is paving the way for the deployment of at-scale smart EV chargers offering new ways to communicate with prosumers and near-real-time interfacing with the EV battery management system. However, due to the shift towards decentralized models and the rapid changes in the demand, the energy market imposes extreme communication requirements as metering and decision processes should be performed at very high frequencies.
Several solutions have been designed to reduce the communication overhead, such as relying on reinforcement learning ~\cite{saputra2019energy} and federated learning ~\cite{taik2020electrical}.
%Work in~\cite{saputra2019energy} adopts a reinforcement learning technique to design an energy demand learning model for electrical vehicle networks without sharing data between charging stations and charging station providers, hence preserving the data privacy and decreasing the communication overhead.

%https://ieeexplore.ieee.org/stamp/stamp.jsp?tp=&arnumber=8752482
%Wireless communication in SG: 
%https://ieeexplore.ieee.org/stamp/stamp.jsp?tp=&arnumber=9023471
%https://www.sciencedirect.com/science/article/pii/S0306261919316599

\vspace{0.2cm}
\textbf{Prosumer Group Formation:} 
The formation of stable PCGs is essential for sustainable energy sharing~\cite{gensollen_stability_2018,rathnayaka_methodology_2014} {\boubakr \cite{ali_synergychain_2021}}. Therefore, it is essential to design strategies for optimal formation, growth, and management of PCGs. The first approach towards the formation of prosumer groups that can be considered is geographical proximity. As the trading market evolves, the similarity of prosumers' energy-sharing behaviors can be used to form stable coalitions. Additionally, the efficiency of PCGs can be further enhanced through unified goals, such as fulfilling the energy demands for external customers, maximizing profits, and reducing costs {\rev \cite{rathnayaka_methodology_2014}}.

%Trust and local reputation of prosumers can be used to study prosumer behavior and form cohesive prosumer groups. 
%P2P, VPP( Virtual Power Plant ), microgrids
%https://ieeexplore.ieee.org/stamp/stamp.jsp?tp=&arnumber=7479542

\begin{figure*}[!t]
	\centering
	\includegraphics[width=0.8\linewidth]{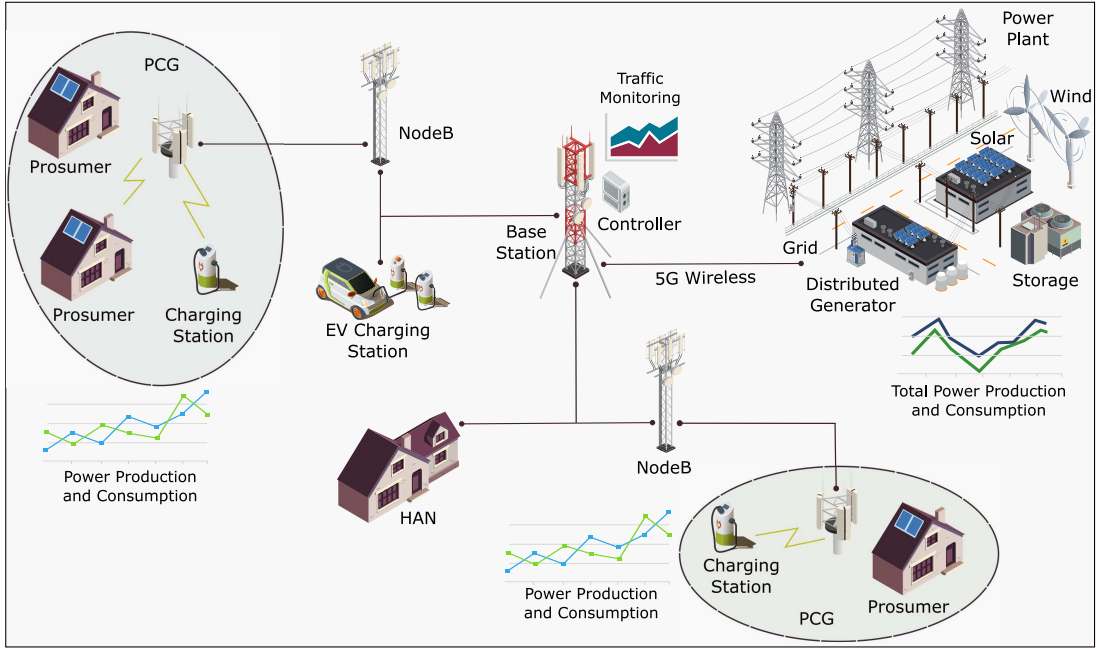}
	\caption{{\rev Prosumers in smart grid: overview and use-cases.}}
	\label{fig:use-cases}
\end{figure*}

\vspace{0.2cm}
\textbf{User \& Data Privacy:}
One of the critical operations in energy market planning is knowledge extraction from historical consumption data. Nonetheless, these data contain sensitive information about users (\eg device usage, household occupancy), which imposes new requirements related to data privacy and necessitates the design of knowledge extraction methods immune to malicious interception and misuse. The centralized management paradigm is inadequate for the privacy-sensitive power distribution market. In contrast, the decentralized structure for the energy market has become an essential subject in smart grid literature. Additional techniques, such as Battery-based Data Masking, authentication and authorization, and Federated Learning, have been used in this setting to enhance privacy~\cite{triantafyllou_challenges_2020}.

\vspace{0.2cm}
An inadequately planned prosumer market and poorly designed decision processes will severely impact consumer empowerment and the sustainability of energy markets. It is, therefore, crucial to take into account different issues when designing architectures and mechanisms for energy markets.

\section{Smart Grid Prosumer Community Empowered Federated Edge Learning}
\label{sec:design}

\begin{figure*}[!t]
	\centering
	\includegraphics[width=0.9\linewidth]{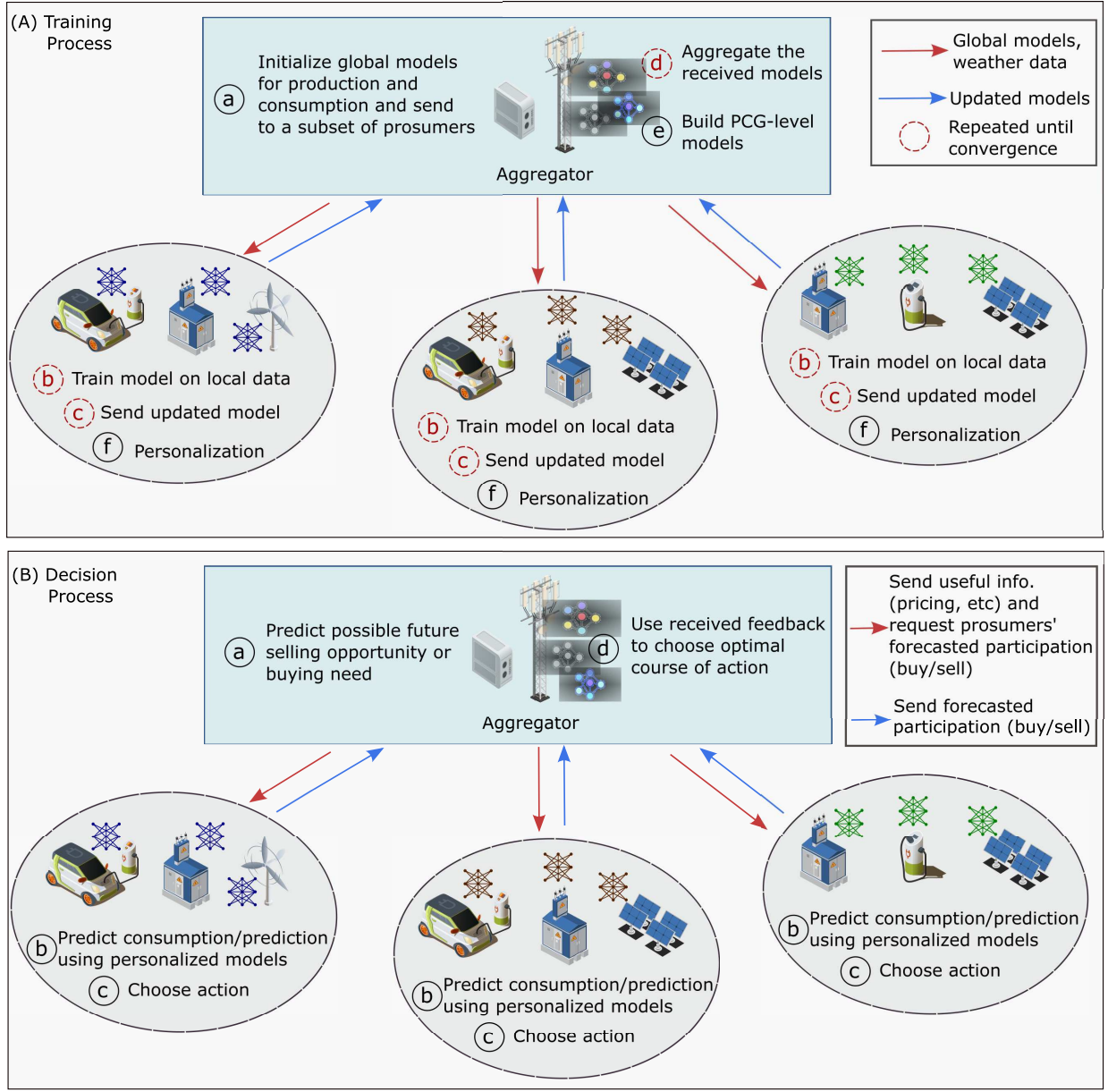}
	\caption{{\rev Training and decision processes of the proposed scheme.}}
	\label{fig:phases}
\end{figure*}

\subsection{System Architecture}
{\boubakr Figure~\ref{fig:use-cases} illustrates a reference architecture for a smart grid. A smart grid is mainly built upon a bi-directional communication between users and utilities. The underlying connectivity between smart grid elements is ensured by a wireless 5G network via base stations and gNodeB. A centralized controller (\eg at the gNodeB level) is employed for traffic monitoring.

To ensure a sustainable energy infrastructure, the grid utilizes a Distributed Generator (DG) and incorporates heterogeneous DERs and produces a time-varying capacity. The integration of DERs requires controlling and monitoring, which can be achieved through computing and communication capabilities of edge and local devices. For instance, smart buildings integrate an agent CPU connected to HAN's sensors/actuators which allows a seamless management of the appliances. 

A set of prosumers within the same administrative domain can collectively form a PCG. A PCG aggregates heterogeneous DERs' capacities and enables energy sharing among prosumers and with external entities, either for profit or for free {\rev \cite{rathnayaka_methodology_2014}}. However, due to the heterogeneity of the DERs, price fluctuations, the automation of energy sharing is rather challenging and requires a high level of coordination among different entities. As a result, information (\eg weather data, pricing data) and commands flow between the utility, the controller, and prosumers.} 

{\boubakr The main objective is to efficiently utilize each energy source.} For example, during the day, PV can be considered the primary source of energy, while batteries, more specifically the V2G model, can help regulate the needs during the day and be the most relied on at night. In doing so, a PCG utilizes forecasted consumption and production values to determine whether the production goals can be met and make trading decisions accordingly. 

In this article, we design a multi-stage framework for collaborative and sustainable energy sharing for PCGs. 

%The two aspects we consider are sustainable participation through decentralized decisions using through the use of FEEL to train the forecast models. 
{\afaf Figure~\ref{fig:phases} shows in detail the two processes of the proposed framework: Training and Decision. The training is performed using FEEL, which is mainly privacy-preserving collaborative training of the load forecasting models. While in the decision process, individual prosumers use the resulting models to make predictions to make optimal decisions.}

\subsection{{\afaf Training Process: FEEL-based Short-term Load Forecasting}}
Current energy trading markets use one-day ahead predictions. However, these predictions are often inaccurate, which in consequence leads to unmatched energy production. By leveraging the HAN edge equipment and the rapid 5G communication \cite{abouaomar_resource_2021,filali_multi-access_2020}, decisions can be taken in near-real-time to match the unpredictable demand~\cite{chen_trading_2019}. {\afaf For instance, EVs charging loads are small and unpredictable in the long-term~\cite{wu_multi-time_2019}, making them inadequate for one-day ahead planning in the V2G context.  In contrast, the solar power production curve over a day is usually bell-shaped and can be easily predicted based on weather data.  Consequently, prediction horizons should be chosen depending on the predictability of each DER type and the required data to make the prediction. Values related to residents' behaviors should have shorter prediction horizons (\eg a few seconds or minutes ahead ) compared to values that depend on the weather (\eg up to several hours ahead).}

For each prosumer, the prediction model takes as an input a series of surplus power data for the last few intervals and produces an estimation of the surplus or shortage power in the next interval. Targeting more accurate predictions, we use separate models for production and consumption to calculate the estimate.
Energy production depends mostly on weather conditions. The required meteorological data for forecasting can be easily obtained by local weather stations and broadcasted by the PCG aggregator. DER power prediction uses models that map historical data to output power through mining the potential rules and relationships of the training data. For instance, the PV production curve is usually bell-shaped with fluctuations during the day depending on solar irradiance, whereas wind energy depends on wind speed and direction. While these factors can be forecasted accurately a few hours ahead, load profiles for EVs and consumption can only be forecasted up to a few minutes ahead due to individual behaviors' uncertainty. 

Moreover, the model should be adapted for each individual prosumer. In fact, the consumption load depends on the individual prosumer's habits, whereas for energy, there are additional factors that affect production, such as the direction solar panels are facing and their position. Due to the scarcity of local data, collaboration among PCG members, in addition to energy sharing, involves a distributed training of the prediction models. FEEL allows training models among the PCG members at the edge of the network while preserving the privacy of the prosumers by keeping data locally~\cite{taik2020electrical}. Training models using FEEL ensures that the obtained model is not overfitting as it is trained on a large dataset. In FL, the training of the model takes several communication rounds before converging. Typically, a global model is initialized by the PCG aggregator and sent to a subset of the prosumers who independently train the model using their local historical data and upload their gradient updates to the PCG aggregator. {\afaf~ The participating subset is ideally selected at random, however, due to the communication bottleneck in wireless edge networks, the selection is based on the wireless channel state, the computing resources, and the amount and quality of the training data.} The received local updates are aggregated by averaging, and a global model is obtained. Afterward, the aggregator sends the global model to a new subset of prosumers, and a new iteration begins where each device computes the gradient updates and uploads it, {\afaf until the model converges}. The final model is then broadcasted to the community. To make the model personalized for each prosumer, the obtained model is retrained locally for each prosumer. {\afaf The FEEL process is repeated periodically to adapt the models to different changes in the prosumers' side (\eg new appliances, different habits) and the external conditions. }

\subsection{{\afaf Multi-stage Decision Process}}
As mentioned above, the PCG leverages communication and energy storage to achieve collective values and goals.  However, individual decisions of prosumers are based first on individual self-sufficiency goals. To maintain PCG stability, the trading decisions should take into account the individual prosumers' decisions. As a result, the decision process for energy trading comprises two distinct levels:  

\vspace{0.2cm}
\textbf{At the aggregator level:} 
Based on the overall predicted values of production and consumption, the aggregator decides the next action regarding energy trade: 
\begin{enumerate*}[(i)]
    \item sell energy if the local production is higher than the overall consumption,
    \item buy energy from external parties if the local production is less than the overall consumption, and/or
    \item request V2G to regulate small fluctuations in the demand.
\end{enumerate*}
Regardless of the preliminary decision, the aggregator requests local decisions made by individual prosumers. 

\vspace{0.2cm}
\textbf{At the individual prosumer level:} 
By using local prediction models, each prosumer estimates the difference between short-term local production and overall consumption. The predicted value serves as a basis for the prosumer's participation or to signal a projected shortage. These values are then sent to the aggregator for finalizing the trading decisions. 
Individual prosumers may develop different decisions processes depending on their goals and preferences. For instance, a prosumer with altruistic values would use the energy infrastructure to share electricity with the community rather than selling it. In contrast, a prosumer with monetary goals would use it to sell electricity rather and gain revenue than share it.

Forecasting the difference between consumption and production has a significant influence over prosumer decisions for a subsequent period, which is indeed a fundamental operation for all decisions.
For production, and for each type of DERs, different factors can be considered in the decision process. For example, each prosumer in V2G may take into account the battery level and estimate the length of time the vehicle will still be charging. 

\section{Numerical Results}
\label{sec:evaluation}
This section first introduces the used dataset and then describes the prediction model and implementation details. Later, we discuss the evaluation results.

\vspace{0.2cm}
\textbf{Dataset:}
This research was conducted using real data from the Pecan Street Inc.~\cite{database}. We used circuit-level electricity use data at 15 minutes intervals for a PCG, with PV generation and EV charging data for a subset of the PCG. The PCG is simulated by a subset of 18 prosumers who have similar properties from this dataset. A subset of 5 prosumers has EV data, and comprises the same kind of houses (detached-family homes), located in the same areas (\ie Austin, Texas). The dataset is composed of records between May 2018 and August 2018, with 15 minutes resolution data for PV and overall consumption. We also used one-minute resolution data of EV consumption over a period of two weeks.  

\vspace{0.2cm}
\textbf{Prediction Models:} 
We used identical models for consumption and solar energy production (\textit{Model 1}), where each has two Long-Short Term Memory (LSTM) hidden layers composed of 128 neurons each. We used the Mean Squared Error as the Loss function and Adam as the optimizer. The models are trained using normalized data, transformed into sliding windows using 48-time steps to predict the next value. For the EVs, we trained a similar model (\textit{Model 2}) with 200 LSTM cells in each layer, using 15 past minutes to predict the next 5 minutes.  

We used 90\% of data for training and 10\% for testing. Each model is trained over 25 rounds and retrained using 8 epochs locally, with a subset of 5 prosumers chosen randomly in each round.  The simulations were conducted on a Laptop with a 2.6 GHz Intel i7 Processor, 16GB of RAM memory, and NVIDIA GeForce RTX 2070 graphic card. We used Tensorflow Federated 0.4.0 with Tensorflow 1.13.1 backend.

\vspace{0.2cm}
\textbf{Results:} 

\begin{figure}[!t]
	\centering
	\includegraphics[width=\linewidth]{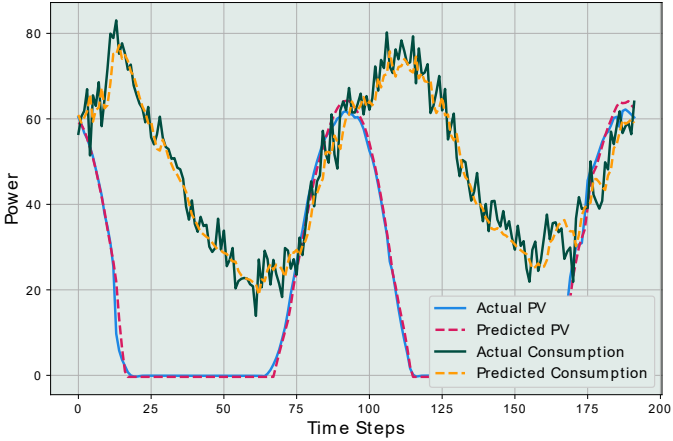}
	\caption{{\boubakr Overall prediction of production and consumption of the PCG at the aggregator level.}}
	\label{fig:overall_prediction}
\end{figure}

\vspace{0.2cm}
\textit{1) Aggregator level models:}
The aggregator has access to the overall consumption of the PCG and the production of the DERs. In our case, it uses the prediction of the consumption alongside the forecasted PV power to make decisions. Figure~\ref{fig:overall_prediction} shows the prediction and the actual power for the PCG for the first 24 hours in the test set.
The Root Mean Square Error (RMSE) of the PV power is negligible $3.99~W$, whereas the overall consumption can be predicted with an RMSE of $4.71~W$. 

\vspace{0.2cm}
\textit{2) Prosumer level models:}
In order to evaluate the models obtained using FEEL, we compared the average RMSE of these models to the average RMSE of a centrally trained model. Table~\ref{tab:3} summarizes the obtained RMSE for different models. In our case, the load forecast is on a granular level (single house) and on a very short term (\ie 15 minutes or less); therefore, the values of RMSE are achieved in Table \ref{tab:3} for various models are reasonable. The error margin is anticipated as similar values have been reported by previous work \cite{taik2020electrical}.

\begin{table}[!b]
	\centering
	\caption{Average Power RMSE for the PCG.}
	\label{tab:3}
	\renewcommand{\arraystretch}{2}
	\begin{tabularx}{.95\linewidth}{|>{\hsize=0.23\hsize}Y|>{\hsize=0.3\hsize}Z|>{\hsize=0.35\hsize}Z|}
		\hline
		\rowcolor[HTML]{bcd9dd}
		~ &
		\textbf{Central Model} &
		\textbf{Personalized Models} \\ \hline
		
		\rowcolor[HTML]{e6e7e9}
		\textbf{PV} &
		$ 0.25 \pm 0.05 $ &
		$ 0.18 \pm 0.04 $ \\\hline
		
		\rowcolor[HTML]{e6e7e9}
		\textbf{Consumption} &
		$ 0.84 \pm 0.28 $ &
		$ 0.65 \pm 0.23 $ \\\hline
		
		\rowcolor[HTML]{e6e7e9}
		\textbf{EV (1min)} &
		$ 0.13 \pm 0.04 $ &
		$ 0.12 \pm 0.04 $ \\\hline

		\rowcolor[HTML]{e6e7e9}
		\textbf{EV (5min)} &
		$ 0.265 \pm 0.04 $ &
		$ 0.277 \pm 0.04 $ \\\hline
	\end{tabularx}
\end{table}

\begin{figure}[!t]
	\centering
	\includegraphics[width=9cm, height=6cm]{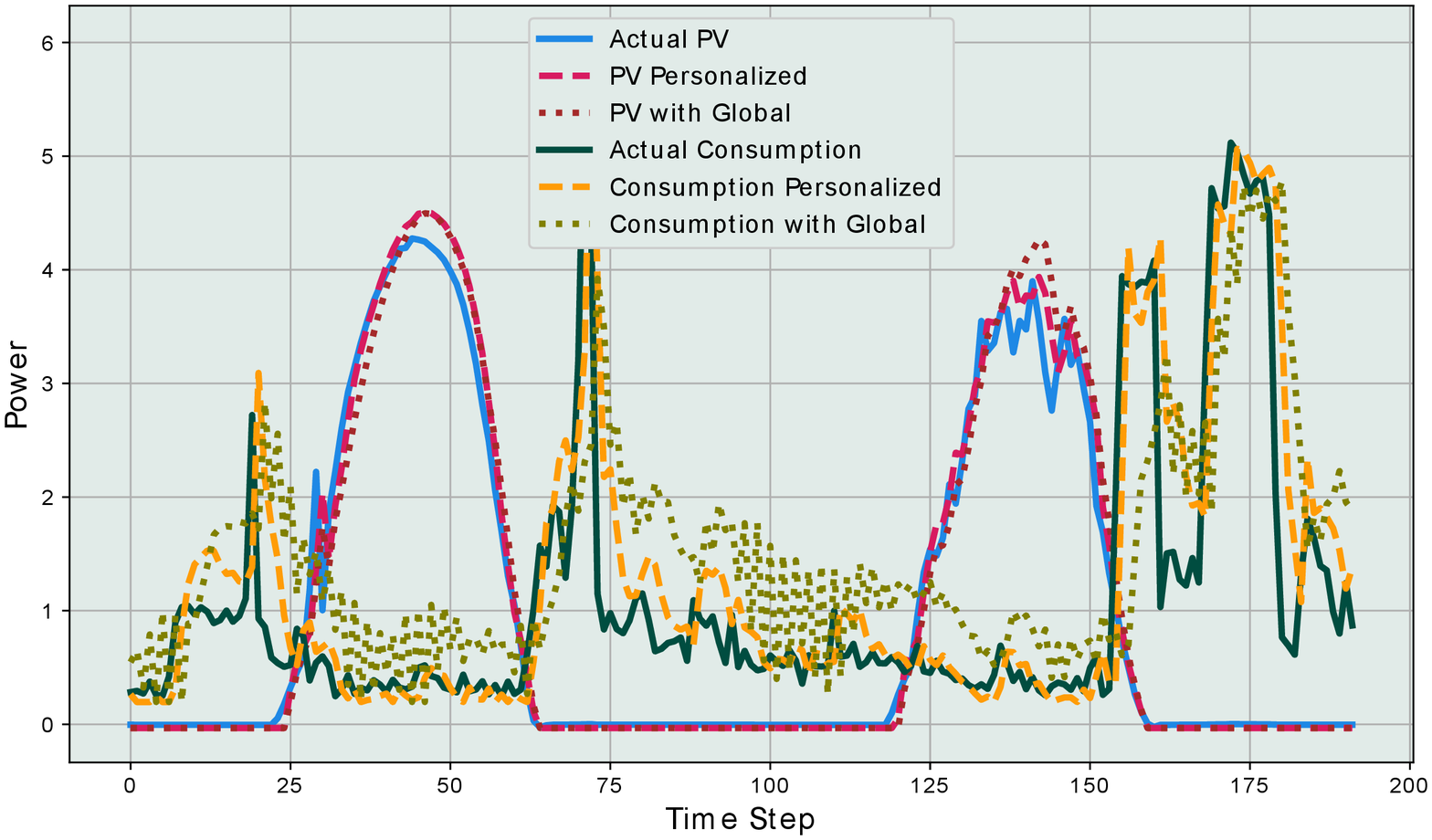}
	\caption{{\boubakr Prediction of production and consumption of the first 24 hours in the test set for a sample prosumer.}}
	\label{fig:sampl_prediction}
\end{figure}

Figure~\ref{fig:sampl_prediction} illustrates the improvements on predictions using personalization with FEEL for a prosumer from the PCG. Both models (PV and consumption) fit the actual data of the prosumer. As energy trading has high accuracy requirements, the improvement in the precision of the prediction will have a significant impact. 

\begin{figure}[!t]
	\centering
	\includegraphics[width=\linewidth]{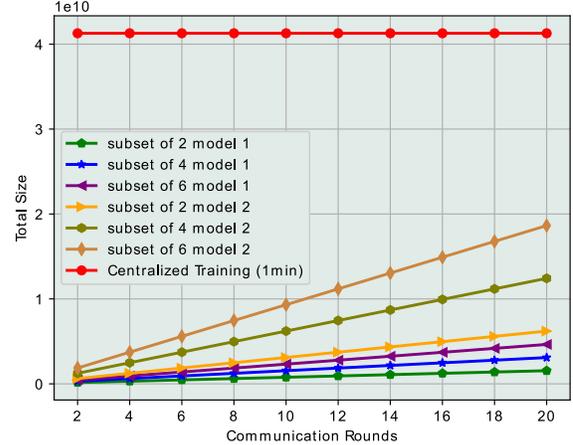}
	\caption{{\boubakr Total size of data exchanged over wireless networks when training models using FEEL vs. centralized training.}}
	\label{fig:communication}
\end{figure}

\vspace{0.2cm}
\textit{3) Communication and Computation Overhead:}
Figure~\ref{fig:communication} shows the total data size sent over wireless networks when using FEEL compared to sending the data in one-minute resolution. While scheduling a larger number of prosumers in each round is preferable, more updates are exchanged over the network leading to costly communication (\eg in terms of bandwidth, delay, and quality of experience). By adopting FEEL, where training is performed at the edge level, data exchange is less costly than frequent data uploads to a central server. With the growth of the need of more granular data, the gain when it comes to communication will be even more significant. Additionally, several compression and partial participation techniques can be explored to further reduce the communication overhead \cite{konevcny2016federated}.

For the computation overhead, our configuration is able to predict the next step for PV and consumption using \textit{Model 1} just by taking an average of $0.37~ms$, while predicting the next 5 steps using \textit{Model 2} for EV power takes an average of $38~ms$. The multi-step prediction is a costly operation and requires deeper and more powerful models. Nonetheless, it remains necessary to maintain a sustainable prosumer participation. 

{\afaf \section{Open Issues and Future Research Directions}
\label{sec:future}
Through this work, we have identified different open issues and interesting research directions that deserve further investigation:

\vspace{0.2cm}
\textbf{Prosumers Regrouping:} Individual prosumer behavior is impacted by the community's organization, norms, and goals. A sustainable PCG can be extended through virtualization techniques to make a match among a wider number of prosumers who share similar goals and preferences. 

\vspace{0.2cm}
\textbf{Peer to Peer trading:} To achieve a fully decentralized collaboration, Peer-to-Peer FL can be leveraged in the context of smart grid and energy markets. This approach removes the single point of failure that can be inherent in a PCG aggregator-based system. Furthermore, blockchain technology \cite{hajar} can be used for reliable transactions~\cite{ali_synergychain_2021}.

\vspace{0.2cm}
\textbf{Privacy and Security:} Although FEEL is designed with privacy in mind, it is still vulnerable to several attacks such as Backdoor attacks and Poisoning attacks \cite{lcn}. It is, therefore, necessary to investigate mechanisms for more reliable and resilient FEEL in smart grid environments. 
}

\section{Conclusion}
\label{sec:conclusion}
Managing prosumers over wireless networks in smart grids is a challenging task that requires proactive decisions and optimal planning. Given the stochastic nature of consumption and production load profiles and the privacy-sensitive aspect of these data, building predictive models for energy trading automation becomes a challenging operation. In this article, we presented the key enablers for integrating prosumers in the energy market. We discussed several challenges and issues concerning communication, privacy, and planning. Keeping these issues in mind, we designed a multi-stage energy forecasting framework using decentralized decisions based on short-term predictions. By leveraging edge equipment, we show that FEEL is a promising solution for tackling privacy challenges related to model training in PCGs.  Furthermore, we proposed the integration of individual prosumers in the energy trading decisions as a key factor for a sustainable PCG.
% Unlike centralized methods, in the designed framework, FEEL uses edge devices to train models hence reducing the communication overhead and the security risks, as well as preserving data privacy. 

%TC:ignore
\section*{Acknowledgments}
The authors would like to thank the Natural Sciences and Engineering Research Council of Canada, for the financial support of this research.

\bibliographystyle{IEEEtran}
\bibliography{Ref}
%TC:endignore
\end{document}